\documentclass[
  aps,
  prl,
  reprint,
  superscriptaddress
]{revtex4-2}

\usepackage{graphicx}
\usepackage{amsmath}
\usepackage{amssymb}

\begin{document}

\title{Intrinsic Matching Frustration in Fluctuating Finite Systems}

\author{Leonid Rubinovich}
\author{Micha Polak}

\affiliation{
Department of Chemistry,
Ben-Gurion University of the Negev,
Beer-Sheva 84105, Israel
}

\begin{abstract}
We formulate intrinsic matching frustration (IMF), a fluctuation-induced, kinetics-independent reduction in the mean capacity permitted by a prescribed matching rule. For complementary one-to-one matching, the instantaneous capacity is set by the minority population, so fluctuations produce a nonzero mean deficit even when the two populations are balanced on average. At finite size, this deficit depends on the full distribution of the population difference and is determined by its variance alone only in the Gaussian limit. Compartmentalization hides matching capacity by preventing cancellation between local imbalances of opposite sign. Fusion releases this hidden capacity monotonically under coarse graining, producing a measurable recovery of product yield following local reaction to completion.
\end{abstract}

\maketitle

Many physical, chemical, and biological processes require constituents to combine according to a prescribed matching rule. In a finite system, the number of realizable matches is determined not by the total population alone but by the population combination compatible with that rule. Fluctuations generate instantaneous departures from the compatible composition, reducing the mean matching capacity even when the matching condition is satisfied on average. Here we develop this general finite-population principle for complementary one-to-one matching between constituents $A$ and $B$.

Let \(N_A\) and \(N_B\) denote the instantaneous populations. Because every match consumes one constituent of each type, the population-limited matching capacity is set by the minority population,
\begin{equation} k(N_A,N_B) = \min(N_A,N_B) = \frac{N_A+N_B-\lvert N_A-N_B\rvert}{2}. \label{eq:matching_capacity} \end{equation}
For a unit-stoichiometric reaction, $k$ is the maximum product population attainable upon reaction to completion. More generally, it bounds the number of pairs actually formed, which may be further restricted by energetics, kinetics, transport, or spatial organization. Equation \eqref{eq:matching_capacity} is an instantaneous counting identity: the absolute population imbalance reduces the matching capacity relative to that of a balanced population with the same total population.

Defining $\Delta=N_A-N_B$ and $\Phi\equiv\langle|\Delta|\rangle/2$, averaging Eq. ~\eqref{eq:matching_capacity} gives 
\begin{equation} \langle k\rangle = \frac{\langle N_A\rangle+\langle N_B\rangle}{2}-\Phi. \label{eq:mean_matching_capacity} \end{equation}
This identity holds for any joint population distribution. For populations balanced on average, $\langle N_A\rangle=\langle N_B\rangle\equiv\bar N$,
\begin{equation} \langle k\rangle=\bar N-\Phi. \label{eq:balanced_matching_capacity} \end{equation}

We refer to this fluctuation-induced, kinetics-independent reduction in the mean matching capacity as intrinsic matching frustration (IMF), and to $\Phi$ as the matching-frustration deficit. The mean unmatched excess is $2\Phi=\langle|\Delta|\rangle$. Here, “frustration” denotes unrealized capacity caused by instantaneous incompatibility between discrete populations, rather than by competing interactions.

Stochastic reaction networks with finite molecular populations are conventionally described by chemical master equations and stochastic trajectories~\cite{Gillespie1977}. In spatially extended $A+B$ annihilation systems, fluctuations in the initial local population imbalance are known to leave majority-species excesses and to produce segregation and anomalous reaction kinetics, particularly when diffusion is inefficient in averaging these fluctuations~\cite{ToussaintWilczek1983,ClementKopelmanSander1991}. Dynamic-compartment descriptions further incorporate stochastic changes in compartment populations~\cite{DusoZechner2020}. Passive droplet encapsulation generates fluctuating molecular copy numbers, often described by Poisson loading statistics~\cite{Collins2015}, while controlled droplet platforms enable on-demand fusion~\cite{Mohan2014,Kaminski2017} and multistep reaction protocols~\cite{Kaminski2017,Nightingale2014}. The present work isolates a distinct, kinetics-independent population-counting constraint underlying such imbalance effects. Before specifying diffusion, reaction rates, or geometry, the instantaneous populations alone bound the number of realizable $A-B$ pairs. We therefore ask: Given the instantaneous populations, what product yield is permitted by population compatibility alone, irrespective of the rate law, transport mechanism, or spatial encounter dynamics?

Although Eq. \eqref{eq:matching_capacity} is elementary, its consequences include finite-size nonuniversality between ensembles with identical first two imbalance moments, emergent variance universality in the Gaussian regime, and an exact, monotonically recoverable product-yield signature under compartment fusion.

\textit{Finite-size distribution dependence.---} For two independent Poisson populations with equal average $\bar N$, the imbalance follows a symmetric Skellam distribution, giving 
\begin{equation} \Phi_{\rm Sk}(\bar N) = \bar N e^{-2\bar N} \left[I_0(2\bar N)+I_1(2\bar N)\right], \label{eq:skellam_mismatch} \end{equation} 
where $I_\nu$ is the modified Bessel function of the first kind of order $\nu$. 
Although this expression follows mathematically from the first absolute moment of the symmetric Skellam distribution, its significance here is physical: it gives the exact finite-population loss of stoichiometric matching capacity for independently fluctuating reactant populations. More generally, Eqs. \eqref{eq:mean_matching_capacity} and \eqref{eq:balanced_matching_capacity} extend this interpretation beyond Poisson statistics to arbitrary joint population distributions. For $\bar N\gg1$, the deficit grows subextensively as $\Phi_{\rm Sk}(\bar N)\sim\sqrt{\bar N/\pi}$, so that its relative magnitude vanishes as $\Phi_{\rm Sk}/\bar N\sim1/\sqrt{\pi\bar N}$. The derivation and higher-order asymptotics are given in Supplemental Material, Sec.~A~\cite{SM}. 

\begin{figure*}[t] \centering \includegraphics[width=\textwidth]{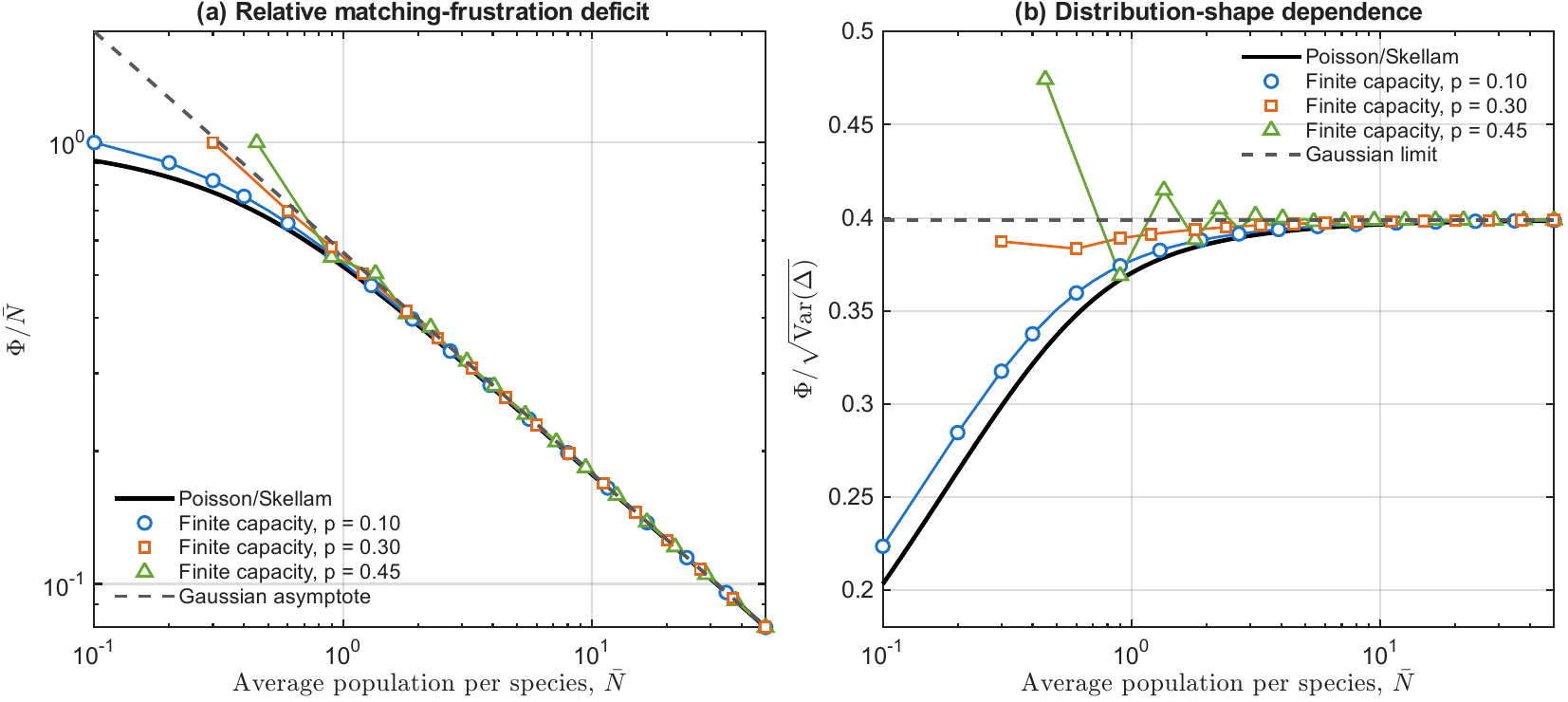} \caption{Intrinsic matching frustration in two finite-population ensembles. (a) Relative deficit $\Phi/\bar N$ for independent Poisson populations and finite-capacity ensembles with $\bar N=Mp$. Here, $\bar N$ is the ensemble-average population of each species and need not be an integer. The dashed line is the Gaussian asymptote $\Phi/\bar N=1/\sqrt{\pi\bar N}$. (b) Variance-normalized deficit. Although both models satisfy $\operatorname{Var}(\Delta)=2\bar N$, they yield different values of $\Phi$ at finite size, showing that the imbalance variance alone does not determine the matching capacity. All results approach the Gaussian limit $\Phi/\sqrt{\operatorname{Var}(\Delta)}=1/\sqrt{2\pi}$. For the finite-capacity model, the indicated values of $p$ are the single-state occupation probabilities for each species, symbols correspond to integer $M=\bar N/p$, and lines connecting them are guides to the eye. The pronounced alternation at larger $p$ is an even--odd finite-size effect: as the vacancy probability $1-2p$ decreases, the imbalance distribution approaches the fixed-occupancy limit in which $\Delta$ has the same parity as $M$.} \label{fig:imf} \end{figure*}

To test whether the imbalance variance determines $\Phi$, we next consider an open finite-capacity population ensemble with $M$ equivalent occupancy states. Each state may be vacant or occupied by one constituent of type $A$ or $B$, with respective probabilities $1-2p$, $p$, and $p$. The equilibrium population distribution is multinomial and satisfies \begin{equation} \langle N_A\rangle=\langle N_B\rangle=Mp\equiv\bar N, \qquad \operatorname{Var}(\Delta)=2Mp=2\bar N. \label{eq:finite_capacity_moments} \end{equation} The imbalance variance is therefore identical to that of the symmetric Skellam distribution at the same $\bar N$. Unlike the unbounded Skellam distribution, however, the finite-capacity imbalance can assume only integer values in the bounded interval $-M\leq\Delta\leq M$. The exact distribution and its generating function are given in Supplemental Material, Sec.~B~\cite{SM}. Exact evaluation gives $\Phi_{M,p}\neq\Phi_{\rm Sk}(\bar N)$ at finite size despite identical average populations and identical imbalance variances, as shown in Fig.~\ref{fig:imf}. This provides an explicit counterexample to a variance-only description of finite-system matching: two ensembles that are indistinguishable at the level of the first two imbalance moments can have different population-limited capacities. Only in the Gaussian limit does the variance become sufficient. General bounds on $\Phi$, including those for bounded imbalance distributions, are given in Supplemental Material, Sec.~C~\cite{SM}.

\textit{Gaussian regime.---} For sufficiently large populations, the standardized imbalance approaches a zero-average Gaussian, for which \begin{equation} \frac{\Phi}{\sqrt{\operatorname{Var}(\Delta)}} \longrightarrow \frac{1}{\sqrt{2\pi}}. \label{eq:normalized_gaussian_mismatch} \end{equation} The derivation is given in Supplemental Material, Sec.~D~\cite{SM}. Equation \eqref{eq:normalized_gaussian_mismatch} is universal within the Gaussian regime: the fluctuation mechanism enters through $\operatorname{Var}(\Delta)$, whereas the normalized deficit is independent of the detailed population statistics. Outside this regime, $\Phi$ generally depends on the full imbalance distribution. Both models approach the Gaussian result in the large-system limit, while the finite-capacity model exhibits increasingly pronounced even--odd corrections as $p$ approaches $1/2$.

\textit{Compartmentalization hides matching capacity.---}We distinguish random loading from subsequent reaction dynamics. Each compartment \(i\) is assigned initial populations \((N_{A,i},N_{B,i})\) drawn from a specified loading distribution. Passive independent encapsulation gives Poisson marginals as an important special case, but our results do not require Poisson statistics. After loading, intercompartment exchange of the reacting species is assumed negligible over the local-reaction timescale. Compartmentalization constrains one-to-one matching because oppositely signed local imbalances cannot cancel before compartments are mixed. 

Consider $C$ elementary compartments, indexed by $i=1,\ldots,C$, with local imbalances $\Delta_i=N_{A,i}-N_{B,i}$. The matching capacity hidden by spatial separation equals the difference of the globally mixed and locally restricted capacities
\begin{align} \delta k &\equiv k_{\rm glob}-k_C \nonumber\\ &= \frac{1}{2} \left[ \sum_{i=1}^{C} \lvert \Delta_i\rvert - \left\lvert\sum_{i=1}^{C} \Delta_i\right\rvert \right] \geq 0. \label{eq:fusion_yield} \end{align} 
This follows from the triangle inequality. Equality holds only when all nonzero local imbalances have the same sign; otherwise, global mixing creates additional matching capacity. The derivation of Eq. \eqref{eq:fusion_yield} is given in Supplemental Material, Sec.~E~\cite{SM}.

\begin{figure*}[t] \centering \includegraphics[width=\textwidth]{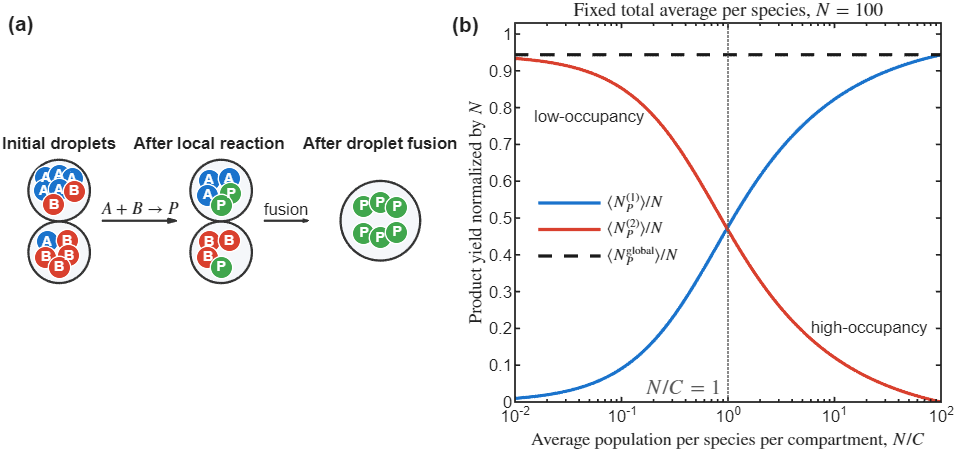} \caption{ Fusion reveals matching capacity hidden by compartmentalization. (a) Local reaction leaves complementary residual excesses in separate droplets; fusion converts them into a second product yield $N_P^{(2)}=k_{\rm glob}-k_C$. (b) Average prefusion, postfusion, and total yields for $C$ equivalent Poisson-loaded droplets at fixed $N=100$. The crossover near $N/C\sim1$ separates the low-occupancy ($N/C\ll1$) and high-occupancy ($N/C\gg1$) regimes, where $N/C$ is the average population of each species per compartment.} \label{fig:fusion} \end{figure*}

\textit{Fusion releases hidden matching capacity.—} Controlled droplet systems can preserve isolated microreactors and subsequently initiate content mixing through droplet manipulation and fusion~\cite{Mohan2014,Kaminski2017}. Related platforms enable controlled multistep reactions in droplet reactors~\cite{Kaminski2017,Nightingale2014}. Isolated droplets therefore provide a direct physical realization. Suppose that irreversible unit-stoichiometric reaction $A+B\to P$ first proceeds to completion within each isolated droplet. The first-stage product is $N_P^{(1)}=k_C$, and each droplet retains only its local excess species. Fusion followed by reaction to completion produces \begin{equation} N_P^{(2)} = \delta k. \label{eq:np2} \end{equation} Thus, the postfusion yield measures exactly the matching capacity hidden by the compartment boundaries. General dynamic-compartment frameworks can represent fusion as a stochastic compartment-level event and subsequently evolve the internal reaction network~\cite{DusoZechner2020}. Equation \eqref{eq:np2} provides a complementary result: under complete local reaction, the \emph{integrated} postfusion product increment is fixed pathwise by the prefusion imbalances and can therefore be obtained without solving the reaction dynamics. Unlike a kinetic modification of a rate constant or relaxation law, this increment depends only on the local population configuration at the time of compartment isolation. It vanishes when all nonzero $\Delta_i$ have the same sign and is positive precisely when fusion permits oppositely signed local imbalances to cancel. Equation \eqref{eq:np2} also provides an operational measure of hidden compartment-to-compartment heterogeneity. For a globally balanced realization, $\sum_i\Delta_i=0$, it reduces to $N_P^{(2)}=\sum_i|\Delta_i|/2$. The integrated postfusion yield therefore measures the total absolute local population imbalance without requiring composition-resolved measurements of the individual compartments. In the fusion protocol of Fig.~\ref{fig:fusion}, this otherwise statistical constraint is converted into a directly measurable product signal.

More generally, progressive fusion produces a monotonic sequence of reaction yields that probes the coarse-graining dependence of hidden local imbalances. Fusion is not the only conceivable mechanism for releasing this capacity. Intercompartment molecular exchange would also allow complementary residual populations to react. Such exchange, however, would consume part of the hidden capacity before fusion and reduce the ensuing secondary yield. The ideal protocol therefore assumes negligible exchange of the reacting species during the local-reaction stage. Fusion then provides a controlled transition from isolated local completion to collective completion, converting the total compartmentalized capacity deficit into directly measurable product.

\textit{Monotonic capacity release under coarse graining.---} Let $\mathcal P$ denote a grouping of the compartments into mutually isolated, disjoint groups $g\in\mathcal P$. The imbalance of group $g$ is
\begin{equation}
\Delta_g=\sum_{i\in g}\Delta_i.
\label{eq:group_imbalance}
\end{equation}
The matching capacity when mixing is permitted within each group, but not between different groups, is
\begin{equation}
k(\mathcal P)=\frac{1}{2}\left[\sum_{i=1}^{C}(N_{A,i}+N_{B,i})-\sum_{g\in\mathcal P}|\Delta_g|\right].
\label{eq:partition_capacity}
\end{equation}
If $\mathcal P'$ is obtained by merging groups $g_1$ and $g_2$, the
triangle inequality gives
\begin{equation}
\begin{aligned}
k(\mathcal P')-k(\mathcal P)
&=\frac12\left(|\Delta_{g_1}|+|\Delta_{g_2}|\right.\\
&\hspace{12mm}\left.-|\Delta_{g_1}+\Delta_{g_2}|\right)
\geq0.
\end{aligned}
\label{eq:S_pairwise_increment}
\end{equation}
Repeated application proves monotonicity for an arbitrary fusion tree. Thus, removing boundaries can only release population-limited matching capacity, both realization by realization and on average.

The Supplemental Material~\cite{SM} gives further statistical examples for equivalent Poisson-loaded and correlated Gaussian compartments, together with extensions to closed finite systems, secondary capture of intact residual populations, parity-constrained homodimerization, and multicomponent matching.

In conclusion, intrinsic matching frustration is a general, kinetics-independent limitation on one-to-one matching in fluctuating finite populations. Its mean deficit is given by the mean absolute population imbalance and, at finite size, depends on the full imbalance distribution rather than on its variance alone. A universal variance-only description emerges in the Gaussian limit. For fixed realized local populations, compartment boundaries do not generate the local imbalances but prevent cancellation between imbalances of opposite sign, thereby hiding globally available matching capacity. For a unit-stoichiometric reaction completed before and after fusion, progressive fusion releases this capacity monotonically under coarse graining, converting the released capacity exactly into additional product and making its coarse-graining dependence directly measurable.

\clearpage
\onecolumngrid
\begingroup
\setcounter{secnumdepth}{1}
\renewcommand{\thesection}{\Alph{section}}
\setcounter{section}{0}
\setcounter{equation}{0}
\renewcommand{\theequation}{S\arabic{equation}}

\begin{center}
{\large\bfseries Supplemental Material for\\[4pt]
``Intrinsic Matching Frustration in Fluctuating Finite Systems''\par}
\vspace{1em}
Leonid Rubinovich and Micha Polak\\
Department of Chemistry, Ben-Gurion University of the Negev,\\
Beer-Sheva 84105, Israel
\end{center}
\vspace{1em}

\section{Exact deficit for independent Poisson populations} \label{sec:skellam} Consider two independent Poisson populations, \begin{equation} N_A,N_B\sim {\rm Poisson}(\bar N), \qquad \langle N_A\rangle=\langle N_B\rangle=\bar N. \label{eq:S_poisson_populations} \end{equation} Their imbalance, \begin{equation} \Delta=N_A-N_B, \label{eq:S_imbalance} \end{equation} follows the symmetric Skellam distribution \begin{equation} P(\Delta=d) = e^{-2\bar N}I_{|d|}(2\bar N), \qquad d\in\mathbb Z, \label{eq:S_skellam_distribution} \end{equation} where $I_n(x)$ is the modified Bessel function of the first kind. Independence gives \begin{equation} \operatorname{Var}(\Delta) = \operatorname{Var}(N_A)+\operatorname{Var}(N_B) = 2\bar N. \label{eq:S_skellam_variance} \end{equation} The matching-frustration deficit is \begin{align} \Phi_{\rm Sk}(\bar N) &= \frac{1}{2}\langle|\Delta|\rangle \nonumber\\ &= e^{-2\bar N}\sum_{d=1}^{\infty} d I_d(2\bar N). \label{eq:S_skellam_phi_sum} \end{align} The modified-Bessel recurrence relation \begin{equation} I_{d-1}(x)-I_{d+1}(x) = \frac{2d}{x}I_d(x) \label{eq:S_bessel_recurrence} \end{equation} implies \begin{equation} \sum_{d=1}^{\infty}dI_d(x) = \frac{x}{2}\left[I_0(x)+I_1(x)\right]. \label{eq:S_bessel_sum} \end{equation} Setting $x=2\bar N$ yields \begin{equation} \boxed{ \Phi_{\rm Sk}(\bar N) = \bar N e^{-2\bar N} \left[I_0(2\bar N)+I_1(2\bar N)\right]. } \label{eq:S_skellam_phi} \end{equation} For large $z$, \begin{equation} I_\nu(z) \sim \frac{e^z}{\sqrt{2\pi z}} \left[ 1-\frac{4\nu^2-1}{8z} +\mathcal O(z^{-2}) \right]. \label{eq:S_bessel_asymptotic} \end{equation} Applying this expansion to Eq.~\eqref{eq:S_skellam_phi} gives \begin{equation} \Phi_{\rm Sk}(\bar N) = \sqrt{\frac{\bar N}{\pi}} \left[ 1-\frac{1}{16\bar N} +\mathcal O(\bar N^{-2}) \right], \label{eq:S_skellam_asymptotic} \end{equation} and hence \begin{equation} \frac{\Phi_{\rm Sk}(\bar N)}{\bar N} = \frac{1}{\sqrt{\pi\bar N}} \left[ 1-\frac{1}{16\bar N} +\mathcal O(\bar N^{-2}) \right]. \label{eq:S_relative_skellam_asymptotic} \end{equation} For small $\bar N$, expansion of Eq.~\eqref{eq:S_skellam_phi} gives \begin{equation} \Phi_{\rm Sk}(\bar N) = \bar N-\bar N^2+\bar N^3 -\frac{5}{6}\bar N^4 +\mathcal O(\bar N^5). \label{eq:S_skellam_small_mean} \end{equation} 

\section{Exact open finite-capacity distribution} \label{sec:finite_capacity} Consider $M$ equivalent capacity states, each of which may be vacant or occupied by a constituent of type $A$ or $B$. In the symmetric equilibrium state, the single-state probabilities are \begin{equation} p_A=p_B=p, \qquad p_V=1-2p, \qquad 0\leq p\leq\frac{1}{2}. \label{eq:S_single_state_probabilities} \end{equation} The equilibrium joint population distribution is multinomial: \begin{equation} P_M(n_A,n_B) = \frac{M!} {n_A!n_B!(M-n_A-n_B)!} p^{n_A+n_B} (1-2p)^{M-n_A-n_B}, \label{eq:S_joint_multinomial} \end{equation} where $n_A,n_B\geq0$ and $n_A+n_B\leq M$. The mean populations are \begin{equation} \langle N_A\rangle = \langle N_B\rangle = Mp \equiv \bar N. \label{eq:S_finite_capacity_means} \end{equation} The imbalance $\Delta=N_A-N_B$ has finite support, $-M\leq\Delta\leq M$. For $d\geq0$, write \begin{equation} N_A=j+d, \qquad N_B=j, \label{eq:S_population_parameterization} \end{equation} where \begin{equation} j=0,1,\ldots, \left\lfloor\frac{M-d}{2}\right\rfloor. \end{equation} Substitution into Eq.~\eqref{eq:S_joint_multinomial} yields \begin{align} P_{M,p}(\Delta=d) ={}& \sum_{j=0}^{\lfloor(M-d)/2\rfloor} \frac{M!} {(j+d)!j!(M-2j-d)!} \nonumber\\ &\times p^{2j+d}(1-2p)^{M-2j-d}. \label{eq:S_finite_capacity_imbalance} \end{align} Symmetry gives \begin{equation} P_{M,p}(\Delta=-d) = P_{M,p}(\Delta=d). \label{eq:S_imbalance_symmetry} \end{equation} The exact matching-frustration deficit is therefore \begin{equation} \Phi_{M,p} = \sum_{d=1}^{M} d\,P_{M,p}(\Delta=d). \label{eq:S_finite_capacity_phi} \end{equation} The same imbalance distribution can be represented compactly by introducing independent single-state variables \begin{equation} X_i= \begin{cases} +1, & \text{with probability }p,\\ -1, & \text{with probability }p,\\ 0, & \text{with probability }1-2p. \end{cases} \end{equation} Since \(\Delta=\sum_{i=1}^{M}X_i\), its generating function is \begin{equation} G_\Delta(z)\equiv\langle z^\Delta\rangle =\left(1-2p+pz+pz^{-1}\right)^M. \end{equation} Equation~\eqref{eq:S_finite_capacity_imbalance} follows by coefficient extraction, \begin{equation} P_{M,p}(\Delta=d)=[z^d]G_\Delta(z). \end{equation} Differentiation at \(z=1\) gives \begin{equation} \langle\Delta\rangle=0,\qquad \operatorname{Var}(\Delta)=2Mp=2\bar N. \end{equation} Thus, the finite-capacity and Skellam models have identical mean populations and imbalance variances at the same \(\bar N\), but generally different deficits at finite size because their full imbalance distributions differ. The finite-support and parity effects are suppressed in the Gaussian limit.

\section{Bounds on the matching-frustration deficit} \label{sec:bounds} Let the imbalance have zero mean and variance \begin{equation} \sigma_\Delta^2 = \operatorname{Var}(\Delta) = \langle\Delta^2\rangle. \end{equation} Because \begin{equation} \Phi=\frac{1}{2}\langle|\Delta|\rangle, \end{equation} the Cauchy--Schwarz inequality gives \begin{equation} \langle|\Delta|\rangle^2 \leq \langle\Delta^2\rangle = \sigma_\Delta^2. \end{equation} Therefore, \begin{equation} \boxed{ \Phi\leq\frac{\sigma_\Delta}{2}. } \label{eq:S_phi_upper_bound} \end{equation} Equality is attained by a symmetric two-point distribution concentrated at $\Delta=\pm\sigma_\Delta$, whenever these values are allowed. For unrestricted support, a fixed nonzero variance does not imply a positive lower bound on $\Phi$. To see this, consider the symmetric distribution \begin{equation} P(\Delta=0)=1-\epsilon, \qquad P\left(\Delta=\pm\frac{\sigma_\Delta}{\sqrt{\epsilon}}\right) = \frac{\epsilon}{2}. \label{eq:S_no_lower_bound_distribution} \end{equation} It has variance $\sigma_\Delta^2$, but \begin{equation} \Phi = \frac{1}{2}\sigma_\Delta\sqrt{\epsilon} \longrightarrow0 \qquad (\epsilon\rightarrow0). \label{eq:S_no_lower_bound} \end{equation} If instead the imbalance has bounded support, \begin{equation} |\Delta|\leq\Delta_{\max}, \end{equation} then pointwise \begin{equation} \Delta^2 \leq \Delta_{\max}|\Delta|. \end{equation} Averaging gives \begin{equation} \sigma_\Delta^2 \leq 2\Delta_{\max}\Phi, \end{equation} and hence \begin{equation} \boxed{ \frac{\sigma_\Delta^2}{2\Delta_{\max}} \leq \Phi \leq \frac{\sigma_\Delta}{2}. } \label{eq:S_finite_support_bounds} \end{equation} The upper bound is attained by a symmetric two-point distribution, whereas the lower bound is attained by a symmetric distribution supported at \(0\) and \(\pm\Delta_{\max}\). Thus, even with bounded support, the variance does not uniquely determine \(\Phi\). 

\section{Gaussian limit} \label{sec:gaussian} For sufficiently large populations, the central limit theorem often implies \begin{equation} \Delta\sim\mathcal N(0,\sigma_\Delta^2), \qquad \sigma_\Delta^2=\operatorname{Var}(\Delta). \label{eq:S_gaussian_imbalance} \end{equation} The mean absolute imbalance is then \begin{align} \langle|\Delta|\rangle &= 2\int_0^\infty \frac{\Delta}{\sqrt{2\pi}\sigma_\Delta} \exp\left( -\frac{\Delta^2}{2\sigma_\Delta^2} \right) \,d\Delta \nonumber\\ &= \sigma_\Delta\sqrt{\frac{2}{\pi}}. \label{eq:S_gaussian_absolute_mean} \end{align} Consequently, \begin{equation} \boxed{ \Phi = \sqrt{\frac{\operatorname{Var}(\Delta)}{2\pi}}, \qquad \frac{\Phi}{\sqrt{\operatorname{Var}(\Delta)}} = \frac{1}{\sqrt{2\pi}}. } \label{eq:S_gaussian_phi} \end{equation} Within the Gaussian regime, the microscopic origin of the fluctuations enters through the imbalance variance alone. This variance-only relation is not generally exact for non-Gaussian imbalance distributions. 

\section{Compartmentalization and the fusion-induced product burst} \label{sec:compartmentalization} Consider $C$ compartments with local populations $N_{A,i}$ and $N_{B,i}$ and local imbalances \begin{equation} \Delta_i=N_{A,i}-N_{B,i}. \label{eq:S_local_imbalance} \end{equation} The locally restricted matching capacity is \begin{align} k_C &= \sum_{i=1}^{C}\min(N_{A,i},N_{B,i}) \nonumber\\ &= \frac{1}{2} \sum_{i=1}^{C} \left( N_{A,i}+N_{B,i}-|\Delta_i| \right). \label{eq:S_local_capacity} \end{align} After global mixing, the matching capacity is \begin{align} k_{\rm glob} &= \min\left( \sum_{i=1}^{C}N_{A,i}, \sum_{i=1}^{C}N_{B,i} \right) \nonumber\\ &= \frac{1}{2} \left[ \sum_{i=1}^{C}(N_{A,i}+N_{B,i}) - \left|\sum_{i=1}^{C}\Delta_i\right| \right]. \label{eq:S_global_capacity} \end{align} Their difference is \begin{equation} k_{\rm glob}-k_C = \frac{1}{2} \left[ \sum_{i=1}^{C}|\Delta_i| - \left|\sum_{i=1}^{C}\Delta_i\right| \right]. \label{eq:S_capacity_difference} \end{equation} The triangle inequality ensures that this difference is nonnegative. Equality holds if and only if all nonzero local imbalances have the same sign. Now let the irreversible reaction $A+B\rightarrow P$ proceed first to completion inside each isolated compartment. Since every reaction event removes one $A$ and one $B$, each $\Delta_i$ is conserved during this stage. The prefusion product population is \begin{equation} N_P^{(1)}=k_C. \label{eq:S_first_stage_product} \end{equation} After local reaction completion, compartment $i$ contains the residual populations \begin{equation} N_{A,i}^{\rm res}=(\Delta_i)_+, \qquad N_{B,i}^{\rm res}=(-\Delta_i)_+, \label{eq:S_residual_populations} \end{equation} where $(x)_+\equiv\max(x,0)$. Following fusion, these residual populations react to completion. The additional product population is \begin{align} N_P^{(2)} &= \min\left[ \sum_{i=1}^{C}(\Delta_i)_+, \sum_{i=1}^{C}(-\Delta_i)_+ \right] \nonumber\\ &= \frac{1}{2} \left[ \sum_{i=1}^{C}|\Delta_i| - \left|\sum_{i=1}^{C}\Delta_i\right| \right] \nonumber\\ &= k_{\rm glob}-k_C. \label{eq:S_second_stage_product} \end{align} Therefore, \begin{equation} N_P^{(1)}+N_P^{(2)}=k_{\rm glob}. \label{eq:S_total_product} \end{equation} For a globally balanced realization, $\sum_i\Delta_i=0$, Eq.~\eqref{eq:S_second_stage_product} reduces to \begin{equation} N_P^{(2)} = \frac{1}{2}\sum_{i=1}^{C}|\Delta_i|. \label{eq:S_balanced_second_stage_product} \end{equation} 

\section{Equivalent Poisson-loaded compartments} Suppose that the compartments are statistically equivalent and that \(N_{A,i}\) and \(N_{B,i}\) are independent Poisson random variables with mean \begin{equation} x=\frac{N}{C}, \label{eq:S_mean_per_compartment} \end{equation} where \(N\) is the total mean population of each species. The mean prefusion product yield is \begin{align} \left\langle N_P^{(1)}\right\rangle &= \sum_{i=1}^{C} \left[x-\Phi_{\rm Sk}(x)\right] \nonumber\\ &= N-C\Phi_{\rm Sk}\left(\frac{N}{C}\right). \label{eq:S_mean_first_stage_yield} \end{align} Because sums of independent Poisson random variables are themselves Poisson distributed, the total populations of \(A\) and \(B\) across all compartments are independent Poisson random variables with mean \(N\). The mean total product yield after fusion and reaction to completion is therefore \begin{equation} \left\langle N_P^{(1)}+N_P^{(2)}\right\rangle = N-\Phi_{\rm Sk}(N). \label{eq:S_mean_total_yield} \end{equation} Subtracting Eq.~\eqref{eq:S_mean_first_stage_yield} from Eq.~\eqref{eq:S_mean_total_yield} gives the mean postfusion product increment, \begin{equation} \boxed{ \left\langle N_P^{(2)}\right\rangle = C\Phi_{\rm Sk}\left(\frac{N}{C}\right) -\Phi_{\rm Sk}(N) }. \label{eq:S_mean_second_stage_yield} \end{equation} Thus, the postfusion increment is the difference between the sum of the local matching-frustration deficits and the irreducible deficit associated with the globally mixed population. Using \(x=N/C\), the normalized prefusion yield in Eq.~\eqref{eq:S_mean_first_stage_yield} takes the exact crossover form \begin{equation} \frac{\left\langle N_P^{(1)}\right\rangle}{N} = 1-\frac{\Phi_{\rm Sk}(x)}{x} = 1-e^{-2x}\left[I_0(2x)+I_1(2x)\right]. \label{eq:S_normalized_prefusion_yield} \end{equation} Substitution of the small- and large-\(x\) expansions of \(\Phi_{\rm Sk}(x)\) gives \begin{equation} \frac{\left\langle N_P^{(1)}\right\rangle}{N} = \begin{cases} x-x^2+\dfrac{5}{6}x^3+\mathcal O(x^4), & x\ll1,\\[7pt] 1-\dfrac{1}{\sqrt{\pi x}} +\dfrac{1}{16\sqrt{\pi}\,x^{3/2}} +\mathcal O(x^{-5/2}), & x\gg1. \end{cases} \label{eq:S_prefusion_yield_asymptotics} \end{equation} For \(x\ll1\), simultaneous occupancy by both species is rare and the prefusion reaction is pair-starved. For \(x\gg1\), the relative yield approaches unity with leading deficit \(1/\sqrt{\pi x}=\sqrt{C/(\pi N)}\). Subdivision therefore transfers yield from the prefusion to the postfusion stage without changing the globally available capacity.

\section{Correlated Gaussian compartments} \label{sec:correlated_gaussian} Averaging Eq.~\eqref{eq:S_second_stage_product} gives \begin{equation} \left\langle N_P^{(2)}\right\rangle = \frac{1}{2} \left[ \sum_{i=1}^{C}\left\langle|\Delta_i|\right\rangle - \left\langle|\Delta_{\rm glob}|\right\rangle \right], \qquad \Delta_{\rm glob} = \sum_{i=1}^{C}\Delta_i . \label{eq:S_averaged_fusion_identity} \end{equation} Suppose that the imbalance vector $\boldsymbol{\Delta}=(\Delta_1,\ldots,\Delta_C)$ is multivariate Gaussian, with \begin{equation} \left\langle\Delta_i\right\rangle=0, \qquad \operatorname{Var}(\Delta_i)=\sigma^2. \label{eq:S_gaussian_local_statistics} \end{equation} Because microscopic population imbalances are integer valued, the multivariate Gaussian description should be understood as a large-population or coarse-grained approximation. The total local deficit is then \begin{equation} \Phi_C = \frac{1}{2} \sum_{i=1}^{C} \left\langle|\Delta_i|\right\rangle = \frac{C\sigma}{\sqrt{2\pi}}. \label{eq:S_gaussian_separated_deficit} \end{equation} Moreover, $\Delta_{\rm glob}$ is Gaussian with variance \begin{equation} \sigma_{\rm glob}^2 = \operatorname{Var}(\Delta_{\rm glob}) = \sum_{i,j=1}^{C} \operatorname{Cov}(\Delta_i,\Delta_j). \label{eq:S_general_total_variance} \end{equation} Thus, \begin{equation} \Phi_{\rm glob} = \frac{\sigma_{\rm glob}}{\sqrt{2\pi}}, \end{equation} and \begin{equation} \boxed{ \left\langle N_P^{(2)}\right\rangle = \Phi_C-\Phi_{\rm glob} = \frac{C\sigma-\sigma_{\rm glob}}{\sqrt{2\pi}}. } \label{eq:S_general_correlated_gaussian_burst} \end{equation} Hence, for fixed local variance, correlations affect the Gaussian fusion yield entirely through the variance of the global imbalance. \subsection{Exchangeable correlations} \label{subsec:exchangeable_correlations} For an exchangeable covariance structure, \begin{equation} \operatorname{Cov}(\Delta_i,\Delta_j) = \sigma^2 \left[ (1-\rho)\delta_{ij}+\rho \right], \label{eq:S_exchangeable_gaussian_covariance} \end{equation} the global-imbalance variance is \begin{equation} \sigma_{\rm glob}^2 = C\sigma^2\left[1+(C-1)\rho\right]. \label{eq:S_exchangeable_total_variance} \end{equation} Positive semidefiniteness of the covariance matrix requires \begin{equation} -\frac{1}{C-1}\leq\rho\leq1. \label{eq:S_admissible_correlation_range} \end{equation} Equation~\eqref{eq:S_general_correlated_gaussian_burst} then gives \begin{equation} \boxed{ \left\langle N_P^{(2)}\right\rangle = \frac{\sigma}{\sqrt{2\pi}} \left[ C- \sqrt{C\left\{1+(C-1)\rho\right\}} \right]. } \label{eq:S_exchangeable_gaussian_burst} \end{equation} For $\rho=0$, \begin{equation} \left\langle N_P^{(2)}\right\rangle = \frac{\sigma}{\sqrt{2\pi}} \left(C-\sqrt{C}\right). \end{equation} At $\rho=1$, the imbalances are perfectly aligned and the secondary yield vanishes. At $\rho=-1/(C-1)$, the global imbalance vanishes almost surely and the yield reaches \begin{equation} \left\langle N_P^{(2)}\right\rangle = \frac{C\sigma}{\sqrt{2\pi}}. \end{equation} \subsection{One-dimensional correlations} \label{subsec:correlated_1d} Consider $C$ compartments arranged in a finite one-dimensional chain with open boundary conditions and covariance \begin{equation} \operatorname{Cov}(\Delta_i,\Delta_j) = \sigma_0^2\rho^{|i-j|}, \qquad |\rho|<1. \label{eq:S_exponential_chain_covariance} \end{equation} Here, ``open'' refers to the chain boundaries, not to particle exchange with an external reservoir. Positive $\rho$ favors aligned neighboring imbalances, whereas negative $\rho$ favors oppositely signed imbalances. For a separation $r$, there are $C-r$ compartment pairs. Therefore, \begin{align} \sigma_{\rm glob}^2 &= \sigma_0^2 \left[ C+2\sum_{r=1}^{C-1}(C-r)\rho^r \right] \nonumber\\ &= \sigma_0^2 \left[ C\frac{1+\rho}{1-\rho} - \frac{2\rho(1-\rho^C)}{(1-\rho)^2} \right]. \label{eq:S_variance_correlated_chain} \end{align} The mean secondary yield follows from Eq.~\eqref{eq:S_general_correlated_gaussian_burst} with $\sigma=\sigma_0$: \begin{equation} \boxed{ \left\langle N_P^{(2)}\right\rangle = \frac{C\sigma_0-\sigma_{\rm glob}}{\sqrt{2\pi}}. } \label{eq:S_correlated_chain_burst} \end{equation} For $C\gg1$ at fixed $|\rho|<1$, \begin{equation} \sigma_{\rm glob}^2 = C\sigma_0^2\frac{1+\rho}{1-\rho} +\mathcal O(1), \label{eq:S_correlated_chain_variance_asymptotic} \end{equation} and hence \begin{equation} \frac{\Phi_{\rm glob}}{\Phi_C} \simeq \sqrt{ \frac{1+\rho}{C(1-\rho)} }. \label{eq:S_correlated_deficit_ratio} \end{equation} Positive correlations preserve a larger global imbalance and suppress the secondary yield, whereas anticorrelations enhance cancellation and increase the yield. The nonnegativity of the yield follows generally from Eq.~\eqref{eq:S_second_stage_product}; Gaussianity is needed only to express the absolute moments through the corresponding variances.

\section{Closed finite-system illustration} \label{sec:closed_system} Intrinsic matching frustration does not require reservoir exchange. Consider a closed system with $N_A+N_B=M$, in which each constituent is independently assigned to species $A$ or $B$ with probability $1/2$. Thus, \begin{equation} P(N_A=n)=2^{-M}\binom{M}{n}, \qquad \Delta=N_A-N_B=2N_A-M. \label{eq:S_closed_binomial_distribution} \end{equation} The exact deficit and mean matching capacity are \begin{equation} \Phi_{\mathrm{cl}} = 2^{-M-1} \sum_{n=0}^{M} |2n-M|\binom{M}{n}, \qquad \langle k\rangle = \frac{M}{2}-\Phi_{\mathrm{cl}}. \label{eq:S_closed_deficit} \end{equation} Since \(\operatorname{Var}(\Delta)=M\), the Gaussian limit gives \begin{equation} \Phi_{\rm cl}=\sqrt{\frac{M}{2\pi}}+O(M^{-1/2}), \qquad \langle k\rangle=\frac{M}{2} -\sqrt{\frac{M}{2\pi}}+O(M^{-1/2}), \end{equation} demonstrating that composition fluctuations produce a nonzero deficit even in a closed system with fixed total population. 

\section{Secondary capture of the intact residual population} After completion of \(A+B\to P\), a realization with \(\Delta=N_A-N_B\) contains \begin{equation} N_A^{\rm res}=(\Delta)_+,\qquad N_B^{\rm res}=(-\Delta)_+, \end{equation} so the total residual population is \(|\Delta|\). If \(N_X\) units of a third constituent capture either residual species, the secondary yield is \begin{equation} N_Q=\min(N_X,|\Delta|). \end{equation} In the abundant-capture limit, with both sequential reactions proceeding to completion, \(N_Q=|\Delta|\) and hence $\langle N_Q\rangle=2\Phi$. The realization-wise conservation relation is \begin{equation} \qquad 2N_P+N_Q=N_A+N_B. \end{equation} For balanced mean populations this implies \begin{equation} \langle N_P\rangle+\frac{1}{2}\langle N_Q\rangle=\bar N. \end{equation} These relations assume fixed populations during the sequential reactions; concurrent exchange generally requires a kinetic description.

\section{Parity-constrained homodimerization} \label{sec:homodimerization} For \(A+A\to P\), an instantaneous population \(N_A\) permits \begin{equation} k_{\mathrm{hom}}(N_A) = \left\lfloor\frac{N_A}{2}\right\rfloor = \frac{N_A-(N_A\bmod 2)}{2}. \end{equation} Relative to the continuous stoichiometric capacity $\langle N_A\rangle/2$, the parity-induced deficit is \begin{equation} \Phi_{\mathrm{hom}} \equiv \frac{\langle N_A\rangle}{2} - \langle k_{\mathrm{hom}}\rangle = \frac{1}{2}\Pr(N_A\ \mathrm{odd}). \end{equation} For \(N_A\sim\operatorname{Poisson}(\lambda)\), with \(\lambda=\langle N_A\rangle\), \begin{equation} \Phi_{\mathrm{hom}} = \frac{1-e^{-2\lambda}}{4}, \qquad \langle k_{\mathrm{hom}}\rangle = \frac{\lambda}{2}-\Phi_{\mathrm{hom}}. \end{equation} For \(C\) isolated compartments, each odd local population leaves one unpaired monomer after reaction completion. If \begin{equation} O=\sum_{i=1}^{C}(N_{A,i}\bmod 2) \end{equation} is the number of odd-population compartments, fusion followed by reaction produces \begin{equation} N_P^{(2)} = \left\lfloor\frac{O}{2}\right\rfloor = k_{\mathrm{glob}}^{\mathrm{hom}} - k_C^{\mathrm{hom}} \geq 0. \end{equation} Thus, fusion converts each pair of residual monomers from odd-population compartments into one additional dimer.

\section{Multicomponent matching frustration} \label{sec:multicomponent_IMF} For the unit-stoichiometric reaction \begin{equation} X_1+\cdots+X_q \longrightarrow P, \end{equation} a realization \(\mathbf{N}=(N_1,\ldots,N_q)\) permits \begin{equation} k^{(q)}(\mathbf{N}) = \min_{\alpha=1,\ldots,q}N_\alpha \end{equation} complete products. Relative to the realization-wise mean population per component, the corresponding matching-frustration deficit is \begin{equation} \phi^{(q)}(\mathbf{N}) = \frac{1}{q}\sum_{\alpha=1}^{q}N_\alpha - \min_{\alpha=1,\ldots,q}N_\alpha \geq 0. \label{eq:multicomponent_deficit} \end{equation} Its ensemble average is \(\Phi^{(q)}=\langle\phi^{(q)}\rangle\). For balanced mean populations, \(\langle N_\alpha\rangle=\bar N\), this gives \begin{equation} \left\langle k^{(q)}\right\rangle = \bar N-\Phi^{(q)}. \end{equation} For \(q=2\), Eq.~\eqref{eq:multicomponent_deficit} reduces to \begin{equation} \phi^{(2)} = \frac{|N_1-N_2|}{2}, \end{equation} recovering the binary deficit used in the main text. For \(C\) isolated compartments, the locally restricted and globally mixed capacities are \begin{equation} k_C^{(q)} = \sum_{i=1}^{C} \min_{\alpha}N_{\alpha,i}, \qquad k_{\mathrm{glob}}^{(q)} = \min_{\alpha} \sum_{i=1}^{C}N_{\alpha,i}. \end{equation} Since $\min_\beta N_{\beta,i}\leq N_{\alpha,i}$ for every $i$ and every $\alpha$, \begin{equation} k_C^{(q)} \leq \min_\alpha\sum_iN_{\alpha,i} = k_{\rm glob}^{(q)}. \end{equation} Thus, complete mixing cannot reduce the multicomponent matching capacity. Unlike the binary case, for $q>2$ the deficit is not generally reducible to the absolute value of a single signed population imbalance.

\endgroup
\bibliography{references}
\end{document}